\documentclass[a4,12pt]{article}

\usepackage{epsf}

\newcommand{\nn}{\nonumber\\}
\newcommand{\bra}[1]{\left<#1 \right|}
\newcommand{\ket}[1]{\left| #1 \right>}
\newcommand{\abs}[1]{\left| #1 \right|}
\newcommand{\dz}[1]{\frac{d #1}{2\pi i}}
\newcommand{\rL}{{\rm L}}
\newcommand{\rR}{{\rm R}}
\newcommand{\Q}{Q_{\rm B}}
\newcommand{\half}{\frac{1}{2}}
\newcommand{\CL}{{C_\rL}}
\newcommand{\CR}{{C_\rR}}
\newcommand{\M}{{\rm m}}
\renewcommand{\thepage}{}
\makeatletter
\@addtoreset{equation}{section}
\renewcommand{\theequation}{\thesection.\@arabic\c@equation}
\makeatother

\begin{document}

\begin{titlepage}

\title{
\vspace*{-5ex}
\hfill{\normalsize hep-th/0202133}\\
\vspace{4ex}
\bf Marginal and Scalar Solutions\vspace{2ex}\\
in Cubic Open String Field Theory
\vspace{5ex}}

\author{Tomohiko {\sc Takahashi}\thanks{E-mail address:
 {tomo@asuka.phys.nara-wu.ac.jp}}
\ and\ Seriko {\sc Tanimoto}\thanks{E-mail address:
 {tanimoto@asuka.phys.nara-wu.ac.jp}}
\vspace{3ex}\\
  {\it Department of Physics, Nara Women's University}\\
  {\it Nara 630-8506, Japan}}
\date{February, 2002}

\maketitle
\vspace{8ex}

\begin{abstract}
\normalsize
\baselineskip=19pt plus 0.2pt minus 0.1pt
We find marginal and scalar solutions in cubic open string field theory
by using left-right splitting properties
of a delta function. The marginal solution represents a marginal
deformation generated by a $U(1)$ current, and it is a generalized
 solution of
the Wilson lines one given by the present authors. The scalar
solution has a well-defined universal Fock space expression, and it is
expressed as a singular gauge transform of the trivial vacuum. The
expanded theory around it is
unable to be connected with the original theory by the string field
redefinition. Errors in hep-th/0112124 are corrected in this paper.
\end{abstract}
\end{titlepage}

\renewcommand{\thepage}{\arabic{page}}
\setcounter{page}{1}
\baselineskip=19pt plus 0.2pt minus 0.1pt

\section{Introduction}

Cubic open string field theory (CSFT) \cite{rf:CSFT} has several
classical solutions corresponding to the tachyon
condensation \cite{rf:KS-tachyon,rf:SZ-tachyon}, tachyon 
lump \cite{rf:MSZ,rf:MJMT} and marginal deformations
\cite{rf:SZ-marginal}. 
Since these solutions have been investigated numerically
\cite{rf:GRSZ-2},  
it is necessary to obtain analytic solutions in order to study exact
properties on classical solutions in CSFT. If an analytic solution of
the tachyon vacuum exists,
it is possible to prove Sen's conjectures on the tachyon condensation
\cite{rf:Sen}, or to construct new polynomial string field theory of
closed strings, or to formulate vacuum string field theory completely
\cite{rf:RSZ,rf:GRSZ}. 

There are some attempts to obtain analytic solutions in CSFT
\cite{rf:KP,rf:Schnabl,rf:TT}. In Ref.~\cite{rf:TT}, a Wilson lines
solution and marginal tachyon lump solution are constructed without
using a level truncation scheme with the Siegel gauge. As a result, it turns
out that a branch cut singularity previously found in a marginal
solution \cite{rf:SZ-marginal} should be gauge artifact because the
analytic solution 
outside the Siegel gauge has no singularity. To construct the
solutions, it is helpful to use the identity string field and
left-right splitting properties of a delta function. 

In this paper, we extend the Wilson lines solution to the ones
representing general marginal deformations generated by a $U(1)$
current.  The splitting
properties of a delta function, which is proved as  a generalization of
the ones
in Ref.~\cite{rf:TT}, have an important role in the construction of
marginal solutions. Due to the splitting properties of a delta function,
some operators described by the left and right halves of a string become 
mutually commutable, and then the algebras of the operators 
are splitting into the left and right parts. Since CSFT is described
by a half string formulation \cite{rf:CSFT}, the splitting algebras are
powerful tools in order to find classical solutions in CSFT. 

We also construct a scalar solution in terms of the splitting properties
of a delta function, by which we find splitting algebras
associated with the BRS current. The scalar solution has interesting
features. It has a universal Fock space expression
\cite{rf:SZ-tachyon,rf:SenUniv},  namely it is
written by the matter Virasoro generators and ghost fields, 
since it is made of
the BRS current, the ghost and the identity string field. In addition, the
expanded theory around it can not be connected 
with the original theory by the string field redefinition. 
These properties are required for the tachyon vacuum solution.

Recently, it was proposed that classical solutions are constructed,
which correspond to deformations of D-brane positions
\cite{rf:Kluson}. However, their solutions have singularity originated
in a delta function, and so their Fock space expressions are
ill-defined. On the other hand, our marginal and scalar solutions have
well-defined Fock space expressions, and the marginal solutions
contain the solution corresponding to deformations of D-brane
positions. 

This paper is organized as follows. In sec. 2, we prove the splitting
properties of a delta function.
In sec. 3, we construct general marginal solutions generated by
a $U(1)$ current.
We construct a scalar solution which represents a non-trivial
background in sec. 4. 
We give summary and discussions in sec. 5. We present some detail
calculations in the appendix.

\section{Splitting Properties of Delta Function}

We define a delta function as follows,
\begin{eqnarray}
\label{Eq:delta}
 \delta(w,w') = \sum_{n=-\infty}^\infty w^{-n}w'^{n-1} 
 = \sum_{n=-\infty}^\infty w'^{-n}w^{n-1},
\end{eqnarray}
where $w$ and $w'$ are complex coordinates within a unit circle. If
$f(w)$ has no pole 
except the origin, the delta function satisfies
\begin{eqnarray}
\label{Eq:delta-2}
 f(w) = \oint_{C_0} \frac{dw'}{2\pi i} \delta(w,w')f(w'),
\end{eqnarray}
where $C_0$ denotes the contour which encircles the origin along the unit
circle. 
\begin{figure}[b]
\epsfxsize = 10cm
\centerline{\epsfbox{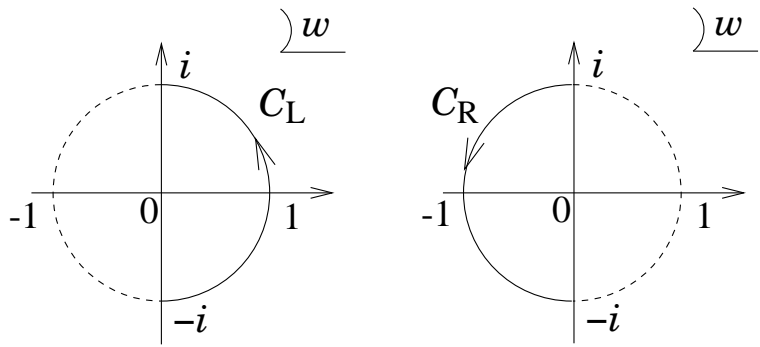}}
\caption{Contours  $C_\rL$ and
 $C_\rR$ along left and right halves of a string.}
\label{fig:fig-LR}
\end{figure}

Let us consider integrations along left and right half unit circles
as depicted in Fig.~\ref{fig:fig-LR}. We find that
\begin{eqnarray}
\label{Eq:CL}
 \int_{C_\rL} \frac{dw'}{2\pi i} w'^n \delta(w,w')
 = \frac{1}{2} w^n + \frac{1}{\pi}
   \sum_{k\neq n}\frac{1}{n-k} w^k \sin\left(
    \frac{(n-k)\pi}{2}\right).
\end{eqnarray}
This implies that the delta function of Eq.~(\ref{Eq:delta}) is not a
delta function if the integration path is the left half only. We perform
the left integration of Eq.~(\ref{Eq:CL}) once more. We find that,
if $m+n\neq 0$,
\begin{eqnarray}
&&\hspace{-3em}\int_{C_\rL} \frac{dw}{2\pi i} w^{m-1}
 \int_{C_\rL} \frac{dw'}{2\pi i} w'^n \delta(w,w') \nn
&=& \frac{1}{\pi}\frac{1}{m+n}\sin\left(\frac{(m+n)\pi}{2}\right)\nn
&&
+\frac{1}{\pi^2} \sum_{k\neq 0,\,m+n}
\frac{1}{k(k-m-n)}\sin\left(\frac{k\pi}{2}\right)
 \sin\left(\frac{(k-m-n)\pi}{2}\right),
\end{eqnarray}
if $m+n=0$,
\begin{eqnarray}
\hspace{-3em}
=\frac{1}{4} +\frac{1}{\pi^2}
\sum_{k\neq 0} \frac{1}{k^2} \sin^2\left(\frac{k\pi}{2}\right).
\end{eqnarray}
Using the formula \cite{rf:TT}
\begin{eqnarray}
 \sum_{k\neq 0,m} \frac{1}{k(k-m)}\sin\left(\frac{k\pi}{2}\right)
 \sin\left(\frac{(k-m)\pi}{2}\right)=\frac{\pi^2}{4}\delta_{m,0},
\end{eqnarray}
we can calculate the infinite series and find the following equation,
\begin{eqnarray}
 \int_{C_\rL}\frac{dw}{2\pi i}w^{m-1} \int_{C_\rL} \frac{dw'}{2\pi i} w'^n
 \delta(w,w') =\int_{C_\rL} \frac{dw}{2\pi i} w^{m+n-1}.
\end{eqnarray}
Therefore, if $f(w)$ and $g(w)$ have no pole except the origin, we find
\begin{eqnarray}
\label{Eq:CLCL}
  \int_{C_\rL}\frac{dw}{2\pi i} \int_{C_\rL} \frac{dw'}{2\pi i}
 f(w) g(w') \delta(w,w') =\int_{C_\rL} \frac{dw}{2\pi i} f(w) g(w).
\end{eqnarray} 
Thus, the delta function of Eq.~(\ref{Eq:delta}) behaves as a delta
function in the double left integrations.
From Eqs.~(\ref{Eq:delta-2}) and (\ref{Eq:CLCL}), other formulas
are given by
\begin{eqnarray}
\label{Eq:CLCR}
&&  \int_{C_\rR}\frac{dw}{2\pi i} \int_{C_\rR} \frac{dw'}{2\pi i}
 f(w) g(w') \delta(w,w') =\int_{C_\rR} \frac{dw}{2\pi i} f(w) g(w),\nn
&&
  \int_{C_\rL}\frac{dw}{2\pi i} \int_{C_\rR} \frac{dw'}{2\pi i}
 f(w) g(w') \delta(w,w') = 0.
\end{eqnarray}
These formulas of Eqs.~(\ref{Eq:CLCL}) and (\ref{Eq:CLCR}) are
generalizations of the ones for the delta function with the Neumann
boundary condition \cite{rf:TT}.

\section{Marginal Solutions}
We consider the ghost field $c(w)$,  and a $U(1)$ current $J(w)$, which
is a dimension one primary field. These fields are expanded by 
\begin{eqnarray}
c(w)= \sum_n c_n w^{-n+1},\ \ \ 
J(w)= \sum_n j_n w^{-n-1}.
\end{eqnarray}
The commutation relations of $j_n$ are given by
\begin{eqnarray}
\label{Eq:jmjn}
[j_m,\,j_n]= m\delta_{m+n}.
\end{eqnarray}
From Eq.~(\ref{Eq:jmjn}), we find the commutation relation of $J(w)$
as
\begin{eqnarray}
 \label{Eq:JJ}
[J(w),\,J(w')]=-\partial_w \delta(w,w').
\end{eqnarray}
Here, we introduce the following operators,
\begin{eqnarray}
&& V_{\rL}(f)=\int_\CL \dz{w} f(w) c\,J(w),\ \ \ 
 V_{\rR}(f)=\int_\CR \dz{w} f(w) c\,J(w),\ \ \ \nn
&&
C_\rL(f) =\int_\CL \dz{w} f(w) c(w),\ \ \ 
C_\rR(f) =\int_\CR \dz{w} f(w) c(w),
\end{eqnarray}
where $f(w)$ is a holomorphic function within the unit circle except the
origin, and satisfies $f(\pm i)=0$. From the splitting formulas of the
delta function, it follows that 
\begin{eqnarray}
\label{Eq:VLVL}
 \left\{V_\rL(f),V_\rL(g)\right\}&=&
  -\int_\CL \dz{w}  \int_\CL \dz{w'} f(w) g(w') c(w) c(w') \partial_w
  \delta(w,w') \nn
 &=& -\int_\CL \dz{w} f(w) g(w) c\partial c(w) \nn
 &=& 
 -\left\{\Q,\,C_\rL(fg)\right\},
\end{eqnarray}
where we have used $f(\pm i)=0$ when we perform the partial
integration. Similarly, we find that
\begin{eqnarray}
\label{Eq:VRVR}
&& \left\{V_\rR(f),V_\rR(g)\right\}=
 -\left\{\Q,\,C_\rR(fg)\right\}, \\
\label{Eq:VLVR}
&& \left\{V_\rL(f),V_\rR(g)\right\}=0.
\end{eqnarray}
Thus, we can obtain left-right splitting algebras in terms of the
splitting property of the delta function.

Let us consider the functions $F_+^{(h)}(w)$ and $F_-^{(h)}(w)$ which, if
we change the coordinate as $w'=-1/w$, transform into
\begin{eqnarray}
 F_+^{(h)}(w') &=& \left(\frac{dw}{dw'}\right)^h F_+^{(h)}(w), \nn
 F_-^{(h)}(w') &=& -\left(\frac{dw}{dw'}\right)^h F_-^{(h)}(w).
\end{eqnarray}
$F_+^{(h)}(w)$ transforms as a dimension $h$ field and $F_-^{(h)}(w)$ has a
opposite sign compared with a dimension $h$ field. Generally, the
functions $F_\pm^{(h)}(w)$ are given by
\begin{eqnarray}
&& F_+^{(h)}(w) = \sum_{n\geq 0} a_n v_n^{(h)}(w), \nn
&& F_-^{(h)}(w) = \sum_{n\geq 0} b_n u_n^{(h)}(w),
\end{eqnarray}
where $v_n^{(h)}$ and $u_n^{(h)}$ are defined by \cite{rf:RZ}
\begin{eqnarray}
u_n^{(h)}(w)=w^{n-h}-(-1)^{n-h}w^{-n-h},\ \ \ 
v_n^{(h)}(w)=w^{n-h}+(-1)^{n-h}w^{-n-h}.
\end{eqnarray}
The $N$-strings vertex of a midpoint interaction is defined by gluing
the boundaries $\abs{w_i}=1\ 
(i=1,2,\cdots, N)$ of $N$ unit disks with the identifications
\cite{rf:LPP1,rf:LPP2}, 
\begin{eqnarray}
 w_i\,w_{i+1} = -1,\ \ \ {\rm for} \abs{w_i}=1,\ {\rm Re}\,w_i\leq 1,
\end{eqnarray}
where $w_{N+1}$ denotes $w_1$. 
If $\phi(w)$ is a dimension $h$ primary field, 
one forms of $dw F_+^{(-h+1)} \phi$ and $dw F_-^{(-h+1)}\phi$
are transformed as
\begin{eqnarray}
&& dw_{i+1} F_+^{(-h+1)}(w_{i+1}) \phi(w_{i+1}) =
dw_i F_+^{(-h+1)}(w_{i}) \phi(w_{i}),\nn
&& dw_{i+1} F_-^{(-h+1)}(w_{i+1}) \phi(w_{i+1}) =
 -dw_i F_-^{(-h+1)}(w_{i}) \phi(w_{i}).
\end{eqnarray}
Then, considering
$c J(w)$ and $c(w)$ as the primary fields, we obtain the following
equations related to star product:
\begin{eqnarray}
\label{Eq:VAB}
&& (V_\rR(F_-^{(1)})A)*B=(-)^{\abs{A}} A*(V_\rL(F_-^{(1)})B), \\
\label{Eq:CAB}
&& (C_\rR(F_+^{(2)})A)*B=-(-)^{\abs{A}} A*(C_\rL(F_+^{(2)})B),
\end{eqnarray}
where $A$ and $B$ are arbitrary string fields, and $\abs{A}$ is $0$ if $A$
is Grassmann even and 1 if it is odd.
Similarly, we find that
\begin{eqnarray}
\label{Eq:VCI}
 V_\rL(F_-^{(1)}) I = V_\rR(F_-^{(1)}) I,
\ \ \ C_\rL(F_+^{(2)}) I = -C_\rR(F_+^{(2)}) I,
\end{eqnarray}
where $I$ denotes the identity string field. The function
$F_+^{(2)}(w)$ must satisfy $F_+^{(2)}(\pm i)=0$, because the ghost
$c(w)$ has 
the midpoint singularity on the identity string field, which is
evaluated by the oscillator expression as \cite{rf:TT}
\begin{eqnarray}
\label{Eq:cI}
 c(w)\ket{I} = \left[-c_0 \frac{w-w^3}{1+w^2}
+c_1 \frac{w^2}{1+w^2} + c_{-1} \frac{1+w^2+w^4}{1+w^2}
+\sum_{n\geq 2} c_{-n} v_n^{(-1)}(w)\right]\ket{I}.
\end{eqnarray}

Now, we obtain a classical solution associated with the $U(1)$ current,
\begin{eqnarray}
\label{Eq:U1solution}
 \Psi_\M = {\rm diag}\left(
a_i V_\rL(F_-^{(1)}) +\half a_i^2\,C_\rL({F_-^{(1)}}^2) \right)I,
\end{eqnarray}
where $a_i$ are parameters and $i$ corresponds to the Chan-Paton
index.
From Eqs.~(\ref{Eq:VLVL}),
(\ref{Eq:VAB}), (\ref{Eq:CAB}) and (\ref{Eq:VCI}), it follows that
\begin{eqnarray}
\Q\Psi_\M &=& \half a_i^2 \{\Q,\,C_\rL({F_-^{(1)}}^2)\} I, \nn
\Psi_\M*\Psi_\M &=& 
  a_i^2 V_\rL(F_-^{(1)})I*V_\rL(F_-^{(1)})I 
  +  \half a_i^2 V_\rL(F_-^{(1)})I*C_\rL({F_-^{(1)}}^2)I \nn
&&
  +   \half a_i^2 C_\rL({F_-^{(1)}}^2)I*V_\rL(F_-^{(1)})I
  +   \frac{1}{4} a_i^4 C_\rL({F_-^{(1)}}^2)I*C_\rL({F_-^{(1)}}^2)I \nn
&=&   \half a_i^2 \{
            V_\rL(F_-^{(1)}),\,V_\rL(F_-^{(1)}) \}I \nn
&=& -\half a_i^2 \{\Q,\,C_\rL({F_-^{(1)}}^2) \} I.
\end{eqnarray}
Then, the classical solution
satisfies the equation of motion, $\Q\Psi_\M + \Psi_\M*\Psi_\M=0$. 
The
solution for $F_-^{(1)}(w)=u_0^{(1)}(w)+u_2^{(1)}(w)$ are give by
Ref.~\cite{rf:TT}. 
Since the classical solution of Eq.~(\ref{Eq:U1solution}) 
dose not refer to
any boundary condition, it is a generalization of the solution
given by Ref.~\cite{rf:TT}.

If we expand the string field around the classical solution, the BRS
charge becomes
\begin{eqnarray}
 \Q' = \Q + a_i V_\rL(F_-^{(1)})-a_j V_\rR(F_-^{(1)})
 +\half a_i^2 C_\rL({F_-^{(1)}}^2) +\half a_j^2 C_\rR({F_-^{(1)}}^2),
\end{eqnarray}
which operates to each $(i,\,j)$-component of the string field
$\Psi_{ij}$. 
Supposed that there is the dimension `zero' field $\varphi(w)$ which
satisfies 
\begin{eqnarray}
\label{Eq:Qphi}
 \left[\varphi(w),\,J(w')\right] = i\,\delta(w,w'),\ \ \ 
 \left[\Q,\,\varphi(w)\right] = -ic\,J(w).
\end{eqnarray}
For the operator $\varphi$, we introduce the left and right integrated
operators,
\begin{eqnarray}
\Phi_\rL(f) = \int_\CL \dz{w} f(w) \varphi(w),\ \ \ 
\Phi_\rR(f) = \int_\CR \dz{w} f(w) \varphi(w).
\end{eqnarray}
From Eq.~(\ref{Eq:Qphi}), it follows that
\begin{eqnarray}
 \left[\Phi_\rL(f),\,\Q\right] = i V_\rL(f),\ \ \ 
 \left[\Phi_\rL(f),\,V_\rL(g)\right] = iC_\rL(fg).
\end{eqnarray}
By these commutation relations, the shifted BRS charge can be written as
\begin{eqnarray}
 \Q'= e^{-B_{ij}(F_-^{(1)})} \Q e^{B_{ij}(F_-^{(1)})},
\end{eqnarray}
where $B$ is defined by
\begin{eqnarray}
 B_{ij}(F_-^{(1)}) = i a_i \Phi_\rL(F_-^{(1)}) -i a_j \Phi_\rR(F_-^{(1)}).
\end{eqnarray}
Therefore, it seems that the shifted theory is transformed into the
original theory by the redefinition of the string field and the classical
solution is trivial. However, it is
true only if the field $\varphi(w)$ exists and the redefinition is
well-defined for the zero-mode part of $\varphi$. For example, we
consider $J\sim i\partial X$ and $\varphi \sim X$. In this case, if the
direction $X$ is compactified, the redefinition is ill-defined
for the zero-mode of $X$. Indeed, the shifted theory 
describes strings in the Wilson lines background, and the classical
solution corresponds to the condensation of the gauge fields
\cite{rf:TT}.

Using the operator $\Phi_\rL$, the classical solution can be rewritten
as
\begin{eqnarray}
 \Psi_\M=\exp(-i\,a_i \Phi_\rL(F_-^{(1)}) I)*\Q\,
        \exp(i\,a_i \Phi_\rL(F_-^{(1)}) I).
\end{eqnarray}
It implies that the classical solution is a gauge transform of
zero string field, namely it is pure gauge. However, as well as the string
field redefinition, there are cases in which the gauge transformation is
ill-defined because of the zero-mode of $\Phi_L$. For the Wilson lines
solution, the classical solution is locally pure gauge, but it is
globally non-trivial configuration, analogously to field theoretical
situations. 

Let us consider the potential height $S[\Psi_\M]$ at the
classical solution. Since the classical solution has parameters $a_i$,
it follows that
\begin{eqnarray}
 \frac{d}{da_i} S[\Psi_\M] = \frac{2}{g}\int (\Q\Psi_\M+\Psi_\M*\Psi_\M)*
\frac{d\Psi_\M}{da_i} = 0.
\end{eqnarray}
Then, we find that $S[\Psi_\M(a_i)]=S[\Psi_\M(a_i=0)]=0$
\cite{rf:KZ}. Thus, the potential height is always zero and $a_i$ correspond
to marginal parameters.

We can construct classical solutions by using $F_+^{(1)}$ instead of
$F_-^{(1)}$. However, $\Phi_\rL(F_+^{(1)})$ can not involve the zero mode
of $\varphi$ because of $v_0^{(1)}(w)=0$. Therefore, the solutions
correspond to pure gauge and physically trivial solutions. 

\section{Scalar Solutions}

The BRS current is defined by
\begin{eqnarray}
 J_{\rm B}(w) = c\left(T_X + \frac{1}{2}T_{\rm gh}\right)(w)
               +\frac{3}{2}\partial^2 c(w),
\end{eqnarray}
where $T_X(w)$ and $T_{\rm gh}(w)$
denote the energy momentum tensors of string coordinates and
reparametrization ghosts,
respectively \cite{rf:KO,rf:FMS}.
The operator product expansions (OPEs) of the BRS current
and the ghost field are given by
\begin{eqnarray}
\label{Eq:JBJB}
 J_{\rm B}(w) J_{\rm B}(w')&=&
  \frac{-(d-18)/2}{(w-w')^3}c\partial c(w')
  +\frac{-(d-18)/4}{(w-w')^2}c\partial^2 c(w')
          -\frac{(d-26)/12}{w-w'} c\partial^3 c(w')+\cdots\nn
       &=& \frac{-4}{(w-w')^3}c\partial c(w')
              +\frac{-2}{(w-w')^2}c\partial^2 c(w')+\cdots,\nn
 J_{\rm B}(w) c(w') &=& \frac{1}{w-w'}c\partial c(w')+\cdots,
\end{eqnarray}
where $d=26$ is the matter central charge of the conformal field theory.
We can expand the BRS current and the ghost field using oscillation
modes,
\begin{eqnarray}
 J_{\rm B}(w)&=& \sum_{n=-\infty}^\infty Q_n w^{-n-1},\nn
 c(w) &=& \sum_{n=-\infty}^\infty c_n w^{-n+1}.
\end{eqnarray}
Since $\{\Q,\,c(w)\}=c\partial c(w)$,
the OPEs of Eq.~(\ref{Eq:JBJB}) can be rewritten in the form of
anti-commutation relations of these oscillators,
\begin{eqnarray}
\label{Eq:QmQn}
 \{Q_m,\,Q_n \} = 2mn\{\Q,\,c_{m+n}\},\ \ \ 
\{Q_m,\,c_n \} = \{\Q,\,c_{m+n} \}.
\end{eqnarray} 
From Eq.~(\ref{Eq:QmQn}), we find the anti-commutation relation of the
BRS current and the ghost,
\begin{eqnarray}
\label{Eq:JBwJBw}
 \{J_{\rm B}(w),\,J_{\rm B}(w')\} &=&
 \left\{\Q,\,2\,\partial_w \partial_{w'} \left(c(w)\delta(w,\,w')\right)
\right\},\nn
\{J_{\rm B}(w),\,c(w')\}
&=& \{\Q,\,c(w)\delta(w,\,w')\}.
\end{eqnarray}

We now define the following operators,
\begin{eqnarray}
&&
 Q_{\rm L}(f)=\int_\CL \dz{w}
f(w) J_{\rm B}(w),
\ \ \ 
 Q_{\rm R}(f)=\int_\CR \dz{w}
f(w) J_{\rm B}(w),
\end{eqnarray}
where $f(w)$ is a holomorphic function within the unit circle except the
origin, and, in addition,  its values at the midpoint is zero, $f(\pm i)=0$. 
From Eq.~(\ref{Eq:JBwJBw}), we can calculate the anti-commutation
relation of the 
operators as follows,
\begin{eqnarray}
\label{Eq:QLQL-derivation}
&&
 \{Q_\rL(f),\,Q_\rL(g)\}\nn
 &=&
\left\{\Q,\,
\int_\CL \dz{w}\int_\CL \dz{w'} f(w)g(w') 
2\,\partial_w \partial_{w'} \left(c(w)\delta(w,\,w')\right)\right\}\nn
&=&
2\left\{\Q,\,
\int_\CL \dz{w}\int_\CL \dz{w'} \partial f(w) \partial g(w') 
c(w)\delta(w,\,w')
\right\},
\end{eqnarray}
where surface terms are vanished due to $f(\pm i)=g(\pm i)=0$.
Using the delta function formula, we find that
\begin{eqnarray}
\label{Eq:QLQL}
 \{Q_\rL(f),\,Q_\rL(g)\}=2\{\Q,\,C_\rL(\partial f \partial g)\}.
\end{eqnarray}
Similarly, other anti-commutation relations are given by
\begin{eqnarray}
\label{Eq:QCLR}
&& \left\{Q_\rR(f),\,Q_\rR(g)\right\}=2\left\{
 \Q,\,C_\rR(\partial f \partial g)\right\}, \nn
&&
\left\{Q_\rL(f),\,C_\rL(g)\right\}=
 \left\{\Q,\,C_\rL(fg)\right\},\nn
&&
\left\{Q_\rR(f),\,C_\rR(g)\right\}=
 \left\{\Q,\,C_\rR(fg)\right\},\nn
&&
\left\{Q_\rL(f),\,Q_\rR(g)\right\}=
\left\{Q_\rL(f),\,C_\rR(g)\right\}=
\left\{Q_\rR(f),\,C_\rL(g)\right\}=0.
\end{eqnarray}
Thus, we can obtain the splitting algebras associated with the BRS
current by using the splitting properties of the delta function.

We consider the properties of the $Q_{\rL(\rR)}$ related to star
product and the identity string field. Since $dw_i F_+^{(0)}(w_i) J_{\rm
B}(w_i)$ is a globally defined one form in the gluing N-strings surface,
we obtain a similar equation to Eqs.~(\ref{Eq:VAB}) and (\ref{Eq:CAB}),
\begin{eqnarray}
\label{Eq:QAB}
 \left(Q_\rR(F_+^{(0)})A\right)*B=-(-)^{\abs{A}} A*\left(
 Q_\rL(F_+^{(0)})B\right).
\end{eqnarray}
Similarly, we find that
\begin{eqnarray}
\label{Eq:QI}
 Q_\rL(F_+^{(0)})I=-Q_\rR(F_+^{(0)}) I.
\end{eqnarray}

Now, we can show that a classical solution is given by
\begin{eqnarray}
&& \Psi_0 = Q_\rL(F_+^{(0)})I + C_\rL(G_+^{(2)})I, 
\end{eqnarray}
where $G_+^{(2)}(w)$ is
\begin{eqnarray}
 G_+^{(2)}(w)=-\frac{\left(\partial F_+^{(0)}(w)\right)^2}{
1+F_+^{(0)}(w)}.
\end{eqnarray}
$G_+^{(2)}(\pm i)$ must be zero in order to cancel the midpoint
singularity of the ghost on the identity, as in the case of the marginal
solutions. 
Indeed, from Eqs.~(\ref{Eq:CAB}), (\ref{Eq:QLQL}), (\ref{Eq:QCLR}),
(\ref{Eq:QAB}) and (\ref{Eq:QI}), we find that the classical equation
satisfies the equation of motion:
\begin{eqnarray}
&&
 \Q\Psi_0 = \{\Q,\,C_\rL(G_+^{(2)})\} I,\nn
&&
\Psi_0*\Psi_0 =
\{\Q,\,C_\rL((\partial F_+^{(0)})^2+F_+^{(0)} G_+^{(2)})\}I.
\end{eqnarray}
Then, it follows that $\Q\Psi_0+\Psi_0*\Psi_0=0$.

If we expand the string field around the classical solution, the shifted
theory has the following BRS charge,
\begin{eqnarray}
 \Q'=\Q+Q(F_+^{(0)})+C(G_+^{(2)}),
\end{eqnarray}
where we define
\begin{eqnarray}
 Q(f)=Q_\rL(f)+Q_\rR(f),\ \ \ C(f)=C_\rL(f)+C_\rR(f).
\end{eqnarray}
If we take $F_+^{(0)}(w)=\exp(h(w))-1$, the classical solution and the
shifted BRS charge are rewritten as
\begin{eqnarray}
&&
\label{Eq:scalsol} 
\Psi_0=Q_\rL(e^h-1) I -C_\rL\left(
(\partial h)^2 e^h \right)I, \\
&&
\label{Eq:shiftedBRS}
\Q'= Q(e^h)-C\left((\partial h)^2 e^h \right).
\end{eqnarray}

Let us consider the redefinition of the string field.
The ghost number currents are given by
\begin{eqnarray}
&&J_{\rm gh}(w) = cb(w),
\end{eqnarray}
where $c(w)$ and $b(w)$ are ghost and anti-ghost fields, respectively.
The OPEs of the ghost current with the BRS
current and with the ghost field are given by
\begin{eqnarray}
\label{Eq:OPEgh1}
\hspace{-1em}
 J_{\rm gh}(w) J_{\rm B}(w') &=&
    \frac{4}{(w-w')^3} c(w')+\frac{2}{(w-w')^2} \partial c(w')
   + \frac{1}{(w-w')^2} J_{\rm B}(w')+\cdots, \\
\label{Eq:OPEgh2}
 J_{\rm gh}(w) c(w') &=&
     \frac{1}{w-w'} c(w')+\cdots.
\end{eqnarray}
We introduce an  operator for
the function $f(w)$ which is holomorphic except the origin,
\begin{eqnarray}
&& q(f) = \oint \frac{dw}{2\pi i} f(w) J_{\rm gh}(w).
\end{eqnarray}
The OPEs of Eqs.~(\ref{Eq:OPEgh1}) and (\ref{Eq:OPEgh2}) give
the following commutation
relations,
\begin{eqnarray}
\label{Eq:comgh1}
&& [q(f),\,Q(g)]
=Q(fg)-2\,C(\partial f \partial g), \\
\label{Eq:comgh2}
&& [q(f),\,C(g)]=C(fg).
\end{eqnarray}
From the commutation relations of Eqs.~(\ref{Eq:comgh1}) and
(\ref{Eq:comgh2}), we find that,
through the transformation generated by $q(f)$,
the BRS charge becomes
\begin{eqnarray}
\label{Eq:redef-Q} 
e^{q(f)}\, \Q\, e^{-q(f)} &=& \Q +[q(f),\,\Q]+\frac{1}{2!}
   [q(f),\,[q(f),\,\Q]]+\cdots \nn
&=& \Q+ Q(f) +\frac{1}{2!}\left\{Q(f^2)
-2C((\partial f)^2)\right\}+\cdots \nn
&=& Q(e^f)- C\left((\partial f)^2 e^f\right).
\end{eqnarray}
Therefore, if the string field $\Psi$ is redefined as $\Psi=e^{q(h)} \Psi'$,
the shifted BRS charge is transformed into the original BRS charge.

In order to interpret the background of the redefined theory,
let us consider the conservation law of $q(h)$ on the
$N$-strings vertex. The gluing $N$-strings surface can be transformed
into the whole complex $z$-plane by the mapping
\begin{eqnarray}
\label{Eq:Nmapping}
 z=e^{\frac{2\pi (1-k) i}{N}}\left(
\frac{1+iw_k}{1-i w_k}\right)^{\frac{2}{N}},
\ \ \ (k=1,\cdots,N).
\end{eqnarray}
Here, $\exp(2\pi (1-k) i/N)\ (k=1,\cdots,N)$ correspond to the $N$
punctures in the $z$-plane, which represent $N$ strings insertions, and
the origin and the infinity in the $z$ plane correspond to the midpoints
of the $N$ strings.
Since $h(w)$ is an analytic scalar and $F_+^{(0)}(\pm i)=0$,
we find that $h(z=0)=h(z=\infty)=0$. Therefore, the conservation law in
the $z$ plane is given by \cite{rf:RZ}
\begin{eqnarray}
 \bra{V_N} \sum_{k=1}^N \int_{C_k} \dz{z} h(z) J_{\rm gh}(z)=0,
\end{eqnarray}
where the contour $C_k$ encircles the puncture at the $k$-string in the
$z$ plane. The anomalous term at the infinity vanishes due to
$h(\infty)=0$.  
We can express the contour integral around the $k$-string's puncture
in terms of the local coordinates $w_k$. Since the transformation low of the
ghost number current $J_{\rm gh}$ is given by
\begin{eqnarray}
\label{Eq:Janomaltrans}
\frac{dz}{dw}\,J_{\rm gh}(z)=J_{\rm gh}(w)
+\frac{3}{2}\,\frac{d^2 z}{d{w}^2} \left(\frac{dz}{dw}\right)^{-1},
\end{eqnarray}
we obtain the following
identity \cite{rf:RZ},
\begin{eqnarray}
\label{Eq:V3Jgh}
&& \bra{V_N}\sum_{k=1}^N q^{(k)}(h)
=\bra{V_N}\sum_{k=1}^N \oint_{C_0} \frac{dw_k}{2\pi i}\, 
 h(w_k)\,J_{\rm gh}(w_k)
= \kappa_N(h)\bra{V_N},\nn
&& \kappa_N(h)=-\frac{3}{2}\sum_{k=1}^N \oint_{C_0} \frac{dw_k}{2\pi i}\,
 h(w_k)\,\frac{d^2 z}{d{w_k}^2} \left(\frac{dz}{dw_k}\right)^{-1},
\end{eqnarray}
where $C_0$ denotes the contour around the origin.
From Eq.~(\ref{Eq:Nmapping}), we find that
\begin{eqnarray}
\label{Eq:kappa}
\kappa_N(h)&=&-\frac{3}{2}\sum_{k=1}^N \oint_{C_0} \frac{dw_k}{2\pi i}\,
 h(w_k)\,\left(\frac{4i}{N}\frac{1}{1+w_k^2}
-\frac{2w_k}{1+w_k^2}\right) \nn
&=&
-6i \oint_{C_0} \frac{dw}{2\pi i}\,
 h(w)\,\frac{1}{1+w^2}
+3N
\oint_{C_0} \frac{dw}{2\pi i}\,
h(w)\,\frac{w}{1+w^2}.
\end{eqnarray}
The shifted action around the classical solution is given by
\begin{eqnarray}
 S=\frac{1}{g}\int \left(\Psi*\Q' \Psi
     +\frac{2}{3}\Psi*\Psi*\Psi\right).
\end{eqnarray}
The action is made of a reflector and a three strings vertex, which are
$2$-strings and $3$-strings vertices, 
respectively. If we redefine the string field as $\Psi=e^{q(h)}\Psi'$,
the action becomes\footnote{In Ref.~\cite{rf:TT2}, the classical
solution was constructed for a  particular function $h(w)$.
However,
$\kappa_2$ was 
missed when it was physically interpreted.
So, the conclusion has changed in
this paper.}
\begin{eqnarray}
\label{Eq:redefaction}
 S=\frac{1}{g}\int \left(e^{\kappa_2(h)}\,\Psi'*\Q \Psi'
     +\frac{2}{3}\,e^{\kappa_3(h)}\,\Psi'*\Psi'*\Psi'\right).
\end{eqnarray}
Supposed that the first term of Eq.~(\ref{Eq:kappa}) has
non-zero value. Since $\kappa_2/2\neq \kappa_3/3$, we can not absorb
these factors by rescaling the string field, and the shifted theory
becomes the theory with 
the original BRS charge and the different coupling constant. So, the
classical solution may represent the dilaton condensation. If the first
term vanishes, the shifted theory become the original one by
rescaling the string field, and so the
classical solution may correspond to a pure gauge solution. However,
each cases should be investigated more 
carefully, because there is a possibility that the redefinition itself
is ill-defined. 

Expanding the ghost current as $J_{\rm gh}(w)=\sum_n q_n w^{-n-1}$, we
can write the operator $q(h)$ as
\begin{eqnarray}
 q(h) &=& q_0 \oint \dz{w} w^{-1}h(w)
   + q^{(+)}(h) + q^{(-)}(h),
\end{eqnarray}
where $q^{(+)}(h)$ and $q^{(-)}(h)$ which correspond to positive and negative
mode parts of $q(h)$ are given by
\begin{eqnarray}
 q^{(+)}(h) &=&
   \sum_{n\geq 1}^\infty q_n \oint \dz{w} w^{-n-1} h(w), \nn
 q^{(-)}(h) &=&
   \sum_{n\geq 1}^\infty q_{-n} \oint \dz{w} w^{n-1} h(w).
\end{eqnarray}
The operator $e^{q(h)}$ is rewritten by the `normal ordered' form as
follows,
\begin{eqnarray}
 e^{q(h)} = \exp\left({q_0 \oint \dz{w}w^{-1}h(w)}\right)
           \exp\left(\frac{1}{2}\left[q^{(+)}(h),\,q^{(-)}(h)\right]\right)
           e^{q^{(-)}(h)} e^{q^{(+)}(h)}.
\end{eqnarray}
If the commutator of $q^{(\pm)}(h)$ has singularity, the operator
$e^{q(h)}$ is ill-defined and we can not redefine the string field by
$\Psi=e^{q(h)}\Psi'$. Then, using such a singular function $h(w)$, we
can obtain a non-trivial classical solution.
From the OPE of the ghost number currents
\begin{eqnarray}
 J_{\rm gh}(w) J_{\rm gh}(w')=\frac{1}{(w-w')^2}+\cdots,
\end{eqnarray} 
the oscillators $q_n$ satisfy $[q_m,\,q_n]=m\delta_{m+n}$. Then, we
can write generally the commutator of $q^{(\pm)}(h)$ as
\begin{eqnarray}
 \left[q^{(+)}(h),\,q^{(-)}(h)\right]
= \oint \dz{w} \oint \dz{w'} \frac{1}{(w-w')^2}h(w)h(w').
\end{eqnarray}

For example, let us consider the classical solution for
\begin{eqnarray}
 h_a(w) &=& \log\left(1+\frac{a}{2}\left(w+\frac{1}{w}\right)^2\right),
\end{eqnarray}
where $a$ is a real parameter which is larger than or equal to $-1/2$.
For this $h_a(w)$, $F_+^{(0)}(w)$ and $G_+^{(2)}(w)$ are given by
\begin{eqnarray}
 F_+^{(0)}(w) &=& e^{h(w)}-1 = \frac{a}{2}\left(
 w+\frac{1}{w}\right)^2 = \frac{a}{2}
\left(v_0^{(0)}(w)+v_2^{(0)}(w)\right), \nn
G_+^{(2)}(w) &=& -(\partial h(w))^2 e^{h(w)}
    =-a^2 w^{-2}\frac{
      \left(w^2-\frac{1}{w^2}\right)^2}{
      1+\frac{a}{2}\left(w+\frac{1}{w}\right)^2}.
\end{eqnarray}
Indeed, we find that $F_+^{(0)}(\pm i)=0$ and $G_+^{(2)}(\pm i)=0$, and
this $h_a(w)$ gives the classical solution by Eq.~(\ref{Eq:scalsol}).
The function $h_a(w)$ has the Laurent expansion as follows (see in
Appendix A),
\begin{eqnarray}
\label{Eq:haexp}
h_a(w)
 &=& -\log(1-Z(a))^2 - \sum_{n=1}^\infty \frac{(-1)^n}{n}
  Z(a)^n \left(w^{2n}+\frac{1}{w^{2n}}\right), \nn
Z(a) &=& \frac{1+a-\sqrt{1+2a}}{a}.
\end{eqnarray}
Using this expansion and Eq.~(\ref{Eq:kappa}), we can evaluate
$\kappa_N(h_a)$ as 
\begin{eqnarray}
 \kappa_N(h_a) &=& -3N \log(1-Z(a)).
\end{eqnarray}
Here, the first term of Eq.~(\ref{Eq:kappa}) vanishes. Therefore, we may
naively expect that the shifted action reduces the original one and the
classical solution might be pure gauge.
 
However, we must consider the string field redefinition in detail as
general discussion.
From Eq.~(\ref{Eq:haexp}), the operator $q(h_a)$ can be
expressed by
\begin{eqnarray}
 q(h_a) &=& -q_0 \log(1-Z(a))^2
+\sum_{n=1}^\infty \frac{(-1)^{n-1}}{n}
(q_{2n}+q_{-2n}) Z(a)^n  \nn
&=&  -q_0 \log(1-Z(a))^2
+q^{(+)}(h_a)+q^{(-)}(h_a),
\end{eqnarray}
where $q^{(+)}$ and $q^{(-)}$ denote the positive and negative modes
part of $q(h_a)$;
\begin{eqnarray}
 q^{(+)}(h_a) = \sum_{n=1}^\infty \frac{(-1)^{n-1}}{n} q_{2n} Z(a)^n,
\ \ \ 
 q^{(-)}(h_a) = \sum_{n=1}^\infty \frac{(-1)^{n-1}}{n} q_{-2n} Z(a)^n.
\end{eqnarray}
Using $[q_m,\,q_n]=m\delta_{m+n}$, we can evaluate the
commutation relation of $q^{(\pm)}$ as follows, 
\begin{eqnarray}
\label{Eq:qpm}
 [q^{(+)}(h_a),\,q^{(-)}(h_a)] = 2\sum_{n=1}^\infty
\frac{1}{n} Z(a)^{2n} = -2\log(1-Z(a)^2).
\end{eqnarray}
Then, we can rewrite the operator $e^{q(h_a)}$
by the `normal ordered' form
\begin{eqnarray}
 e^{q(h_a)} &=& 
\left(1-Z(a)^2\right)^{-1}
\exp\left(-q_0 \log(1-Z(a))^2\right)
e^{q^{(-)}(h_a)} e^{q^{(+)}(h_a)}.
\end{eqnarray}
For $a>-1/2$, this operator is well-defined since $\abs{Z(a)}<1$.
Therefore, the classical solution for $a>-1/2$ should be pure
gauge solution. However, in the case of $Z(a=-1/2)=-1$, this operator
$e^{q(h_a)}$ has singularity and the string field redefinition is
ill-defined. Thus, we can obtain a non-trivial classical solution for
$a=-1/2$. 

In the case of $a=-1/2$, the classical solution is given by
\begin{eqnarray}
\label{Eq:solution}
 \Psi_0 &=& Q_\rL\left(-\frac{1}{4}\left(
   w+\frac{1}{w}\right)^2\right) I
   + C_\rL\left(w^{-2}\left(
   w+\frac{1}{w}\right)^2\right) I.
\end{eqnarray}
Each term of Eq.~(\ref{Eq:solution}) has a well-defined Fock space
expression as follows,
\begin{eqnarray}
&& Q_\rL\left(-\frac{1}{4}\left(
   w+\frac{1}{w}\right)^2\right)\ket{I}
= -\sum_{n=0}^\infty \frac{(-1)^n}{2\pi}
 \left(\frac{2}{2n+1}-\frac{1}{2n+3}
  -\frac{1}{2n-1}\right)Q_{-2n-1}\ket{I}, \nn
&& C_\rL\left(w^{-2}\left(
   w+\frac{1}{w}\right)^2\right)\ket{I} \nn
&&\hspace{.5cm}=\left[\frac{2}{\pi} c_1 + \frac{10}{3\pi} c_{-1}
   +2\sum_{n=1}^\infty \frac{(-1)^n}{\pi}
   \left(\frac{2}{2n+1}
   -\frac{1}{2n+3}-\frac{1}{2n-1}\right)c_{-2n-1}\right]\ket{I},
\end{eqnarray}
where use has been made of Eq.~(\ref{Eq:cI}) and
\begin{eqnarray}
\label{Eq:JBI}
 J_{\rm B}(w)\ket{I}= \sum_{n=1}^\infty Q_{-n}\,v_n^{(1)}(w)\ket{I}.
\end{eqnarray}
If we expand the string field around the classical solution, the shifted
BRS charge is given by
\begin{eqnarray}
 \Q' = \half\Q -\frac{1}{4}\left(Q_2+Q_{-2}\right)
        +2 c_0 + c_2+c_{-2}.
\end{eqnarray}
The BRS current is written by the matter Virasoro generators and the
ghost fields, and the identity string field is given by the Virasoro
generators \cite{rf:RZ}. Therefore, the scalar solution
has a universal expression for arbitrary backgrounds.

Finally, we show that the classical solution for $a=-1/2$ is generated
by a singular gauge transformation.
From the OPE of Eqs.~(\ref{Eq:OPEgh1}) and (\ref{Eq:OPEgh2}), we can
calculate the commutation relations of $J_{\rm gh}$ with $J_{\rm B}$
and c,
\begin{eqnarray}
\label{Eq:JghJB}
&& [J_{\rm gh}(w),\, J_{\rm B}(w')]=J_{\rm B}(w)\delta(w,w')
    -2\partial_w \partial_{w'}\left(c(w)\delta(w,w')\right),\nn
&& [J_{\rm gh}(w),\,c(w')] = c(w)\delta(w,w').
\end{eqnarray} 
We introduce the operators corresponding to the left and right pieces of
$q(h)$, 
\begin{eqnarray}
 q_\rL(f) = \int_{C_\rL}\dz{w} f(w) J_{\rm gh}(w),\ \ \ 
 q_\rR(f) = \int_{C_\rR}\dz{w} f(w) J_{\rm gh}(w),
\end{eqnarray}
where we impose $f(\pm i)=0$. 
From Eq.~(\ref{Eq:JghJB}), these operators satisfy
\begin{eqnarray}
\label{Eq:qLQL}
&& [q_\rL(f),\,Q_\rL(g)]= Q_\rL(fg)-2 C_\rL(\partial f \partial g),\nn
&& [q_\rL(f),\,C_\rL(g)]=C_\rL(fg),
\end{eqnarray}
and the left and right operators are commutable each other.
From Eq.~(\ref{Eq:V3Jgh}), we find as the case of $N=1$
\begin{eqnarray}
\label{Eq:Iq}
 \bra{I}\left(q_\rL (h_a)+q_\rR (h_a)\right)
   =\kappa_1(h_a) \bra{I},\ \ \ \kappa_1(h_a)=-3\log(1-Z(a)).
\end{eqnarray}
From the anomalous transformation law of Eq.~(\ref{Eq:Janomaltrans}), we
find that
\begin{eqnarray}
\label{Eq:Vq}
 \bra{V_N}\left(q_\rL^{(r+1)}(h_a)+q_\rR^{(r)}(h_a)\right)
 = -3\int_{C_\rL}\dz{w}w^{-1} h_a(w)\bra{V_N},
\end{eqnarray}
where $r$ represents the Hilbert space of the $r$-th string.
Using Eq.~(\ref{Eq:haexp}), we can evaluate the coefficient in the right
hand side as 
\begin{eqnarray}
 -3\int_{C_\rL}\dz{w}w^{-1} h_a(w) = -3\log(1-Z(a)) = \kappa_1(h_a).
\end{eqnarray}
From Eqs.~(\ref{Eq:qLQL}), (\ref{Eq:Iq}) and (\ref{Eq:Vq}), the
classical solution of Eq.~(\ref{Eq:scalsol}) for $h_a(w)$ is expressed
as the gauge transform of zero string field,
\begin{eqnarray}
\label{Eq:gaugetransSol}
 \Psi_0 = \exp\left(q_\rL(h_a)I\right)*\Q \exp\left(-q_\rL(h_a)I\right).
\end{eqnarray}

Let us consider the case of $a=-1/2$. Using Eq.~(\ref{Eq:haexp}), we can
write the operator $q_\rL(h_{-1/2})$ by the mode expression,
\begin{eqnarray}
 q_\rL(h_{-1/2}) = -q_0 \log 2 + q_\rL^{(+)}(h_{-1/2})
+ q_\rL^{(-)}(h_{-1/2}),
\end{eqnarray} 
where we define
\begin{eqnarray}
\label{Eq:qLmode1}
 q_\rL^{(+)}(h_{-1/2})&=& -\frac{1}{2}\sum_{n=1}^\infty
  \frac{(-1)^n}{n} q_{2n} \nn
 && -\frac{1}{\pi}\sum_{n=1}^\infty \frac{(-1)^n}{2n-1}
 \left(\beta(n-\frac{1}{2})+\beta(-n+\frac{1}{2})\right) q_{2n-1} \\
\label{Eq:qLmode2}
 q_\rL^{(-)}(h_{-1/2})&=& -\frac{1}{2}\sum_{n=1}^\infty
  \frac{(-1)^n}{n} q_{-2n} \nn
 && -\frac{1}{\pi}\sum_{n=1}^\infty \frac{(-1)^n}{2n-1}
 \left(\beta(n-\frac{1}{2})+\beta(-n+\frac{1}{2})\right) q_{-2n+1}.
\end{eqnarray}
Here, $\beta(z)$ is defined by
\begin{eqnarray}
 \beta(z) = \sum_{n=0}^\infty \frac{(-1)^n}{z+n},
\end{eqnarray}
and it is written by digamma function $\psi(z)=d\log\Gamma(z)/dz$,
\begin{eqnarray}
 \beta(z)= \frac{1}{2}\left(\psi\left(\frac{z+1}{2}\right)
 -\psi\left(\frac{z}{2}\right)\right).
\end{eqnarray}

Using Eqs.~(\ref{Eq:qLmode1}) and (\ref{Eq:qLmode2}), we can calculate
the commutator of $q_\rL^{(\pm)}$ as follows,
\begin{eqnarray}
 [q_\rL^{(+)}(h_{-1/2}),\,q_\rL^{(-)}(h_{-1/2})] &=&
  \frac{1}{2}\sum_{n=1}^\infty \frac{1}{n} \nn
 &&
 +\frac{1}{\pi^2}\sum_{n=1}^\infty
 \frac{1}{2n-1}\left(
  \beta(n-\frac{1}{2})+\beta(-n+\frac{1}{2})\right)^2.
\end{eqnarray}
Thus, the commutator of $q_\rL^{(\pm)}$ has divergence as well as the one of
$q^{(\pm)}$. Rewriting $\exp(\pm q_\rL I)$ as
\begin{eqnarray}
 \exp\left(\pm q_\rL(h_{-1/2}) I\right) &=&
  \exp\left(\pm q_\rL(h_{-1/2})\right) I \nn
  &&\hspace{-2cm}
= 2^{\mp q_0}\exp\left(\frac{1}{2}\left[
      q_\rL^{(+)}(h_{-1/2}),\,q_\rL^{(-)}(h_{-1/2})\right]\right)
      e^{\pm q_\rL^{(-)}(h_{-1/2})}e^{\pm q_\rL^{(+)}(h_{-1/2})},
\end{eqnarray}
we find that the expression of Eq.~(\ref{Eq:gaugetransSol}) has
singularity for $a=-1/2$, and it implies that the classical solution for
$a=-1/2$ can be expressed as a singular gauge transform of the trivial
vacuum.

\section{Summary and Discussion}

In this paper, we proved the splitting properties of the delta function.
Using the splitting properties of the delta function, 
we found the marginal solutions with well-defined Fock space
expressions related to a $U(1)$
current in CSFT. Then, we constructed the scalar solution with a
well-defined universal Fock space expression in CSFT, the theory
expanded around 
which can not be transformed
into the original one by the string field redefinition. In addition, the
scalar solution can be expressed as a singular gauge transform of the
trivial 
vacuum. Though the
non-trivial scalar solution was constructed based on the particular
function $h_{-1/2}(w)$, we can find other solutions by looking for the
function which make the string field redefinition ill-defined. We have
not yet understood whether other solutions connect to each other by gauge
transformations, or each solution corresponds to a different vacuum. 

We expect that the scalar solution
represents the condensation of the tachyon, if various scalar solutions
represent a single vacuum. The reasons are the
following: First, it is impossible to connect the shifted theory to the
original one by the string field redefinition, and so, the classical
solution represents the non-trivial background which is not merely pure
gauge. Secondly, the classical  
solution is scalar and has a universal expression.
Thirdly,
the physical states of the original theory 
are no longer physical in the expanded theory around the classical
solution. Indeed, we can find that $\Q'\ket{\rm phys}\neq 0$ for all
states $\ket{\rm phys}$ such that $\Q \ket{\rm phys}=0$.

Of course, in order to clarify this conjecture, we must prove at least
two propositions. First, there exists no BRS singlet
state for the shifted BRS charge in the Hilbert space
\cite{rf:KugoOjima}, or equivalently, 
the new BRS charge has vanishing cohomology. Secondly, the potential height 
$S[\Psi_0]$ is equal to the D-brane tension\cite{rf:TT3}. At present, we
can not deny 
the possibilities to prove these propositions. It should be noted about
the latter proposition. As in the case of the marginal solutions, we
find that
\begin{eqnarray}
 \frac{d}{da}S[\Psi_0]
  =\int \left(\Q\Psi_0+\Psi_0*\Psi_0\right)*
\frac{d\Psi_0}{da}=0.
\end{eqnarray}
So, the potential height $S[\Psi_0]$ is equal to zero for
$a>-1/2$. However, it may become non-zero value at $a=-1/2$, because the
classical solution is ill-defined for $a<-1/2$ and so $\Psi_0$
is not differentiable at $a=-1/2$.

We should comment on the relation between our scalar solution and a
purely cubic theoretical approach to the tachyon vacuum
\cite{rf:KishiOhmo}. Their argument was based on purely cubic string field
theory (PCSFT) \cite{rf:HLR}. However, the operator of $Q_\rL^2$
appeared in the equation of motion in PCSFT has a midpoint singularity,
which originates in $\delta(\pm i,\,\pm i)$ of the surface term
in Eq.~(\ref{Eq:QLQL-derivation}). Our scalar solution dose not suffer
from the midpoint singularity because such a singular surface term
disappears due to the condition $f(\pm i)=g(\pm i)=0$. If we
find any classical solution in PCSFT, we should formulate it by other
language without midpoint singularity instead of oscillator expansions, 
which is required also for light-cone type string field theories
\cite{rf:KZ}.

\section*{Acknowledgements}
We would like to thank H.~Hata, H.~Itoyama, I.~Kishimoto, T.~Kugo,
K.~Murakami and K.~Ohmori for 
valuable discussions and comments.

\vspace{5ex}
\centerline{\Large\bf Appendix}
\appendix

\section{Laurent Expansion of $h_a(w)$}

First, we find the following equation,
\begin{eqnarray}
\frac{\sin x}{\cosh A-\cos x}
= 2\sum_{n=1}^\infty e^{-nA}\sin(nx)\ \ \ \ (A>0).
\end{eqnarray}
Integrating this equation, we obtain
\begin{eqnarray}
\label{Eq:sikiA}
 \log(\cosh A- \cos x)=A-\log 2
-2\sum_{n=1}^\infty \frac{e^{-nA}}{n}\cos(nx)\ \ \ \ (A>0),
\end{eqnarray}
where the integral constant is given by comparing the values at
$x=0$ or $\pi/2$. This equation is valid also for the case of $A=0$.

If $a>0$, we can rewrite $h_a(w)$ as
\begin{eqnarray}
\label{Eq:xi1}
 h_a(w) &=& \log\left(
           1+\frac{a}{2}\left(w+\frac{1}{w}\right)^2\right) \nn
             &=&  \log(1+2a \cos^2\sigma) \nn
               &=& \log a + \log\left[\left(1+\frac{1}{a}
                \right)
    -\cos(\pi-2\sigma)\right],
\end{eqnarray}
where we take $w=\exp(i\sigma)$.
Applying Eq.~(\ref{Eq:sikiA}) to the second term of Eq.~(\ref{Eq:xi1})
by taking $A=\log[(1+a+\sqrt{1+2a})/a]$, we find that
\begin{eqnarray}
\label{Eq:xi2}
 h_a(w) &=& \log\frac{1+a+\sqrt{1+2a}}{2}\nn
&&-2 \sum_{n=1}^\infty \frac{(-1)^n}{n}
\left(\frac{1+a-\sqrt{1+2a}}{a}\right)^n \cos(2n\sigma).
\end{eqnarray}
If $0>a\geq -1/2$, $h_a(w)$ can be rewritten by
\begin{eqnarray}
 h_a(w) = \log(-a)+\log\left[
 \left(-1-\frac{1}{a}\right)-\cos(2\sigma)\right].
\end{eqnarray}
Applying Eq.~(\ref{Eq:sikiA}) by taking
$A=\log[(1+a+\sqrt{1+2a})/(-a)]$, we have the same expansion form of
Eq.~(\ref{Eq:xi2}) as the case of $a>0$. 

Taking the limit $a\rightarrow 0$ in the right hand-side of
Eq.~(\ref{Eq:xi2}), the result becomes zero and we find that
Eq.~(\ref{Eq:xi2}) holds in the case of $a=0$. Thus, we derive
the formula of Eq.~(\ref{Eq:haexp}).

\end{document}